\documentclass[%
 aip,
 amsmath,amssymb,
 reprint,%
]{revtex4-1}

\usepackage{graphicx}
\usepackage{dcolumn}
\usepackage{bm}
\usepackage{tablefootnote}

\usepackage{lineno}
\usepackage{amsmath}
\usepackage{esvect}
\usepackage{multirow}

\usepackage[utf8]{inputenc}
\usepackage[T1]{fontenc}
\usepackage{mathptmx}
\usepackage{etoolbox}
\usepackage{float}

\makeatletter
\def\@email#1#2{%
 \endgroup
 \patchcmd{\titleblock@produce}
  {\frontmatter@RRAPformat}
  {\frontmatter@RRAPformat{\produce@RRAP{*#1\href{mailto:#2}{#2}}}\frontmatter@RRAPformat}
  {}{}
}%
\makeatother
\begin{document}

\preprint{AIP/123-QED}

\title{Effect of front surface engineering on high energy electron, X-ray and heavy ion generation from Relativistic laser interaction with thick high-Z targets}
\author{J. Twardowski}
 \altaffiliation{Author to whom correspondence should be addressed: \\ twardowski.justin@gmail.com}
 \affiliation{Department of Materials Science and Engineering, The Ohio State University, Columbus, OH 43210, USA.}
\author{C. Kuz}
 \affiliation{Department of Physics, The Ohio State University, Columbus, OH 43210, USA.}
\author{A. S. Bogale}
 \affiliation{Los Alamos National Laboratory, Los Alamos, NM 87545, USA.}
 \affiliation{Center for Energy Research, University of California San Diego, La Jolla, CA 92093, USA.}
 \author{Z. Su}
 \affiliation{Department of Materials Science and Engineering, The Ohio State University, Columbus, OH 43210, USA.}
 \author{A. Lee}
 \affiliation{Department of Physics, The Ohio State University, Columbus, OH 43210, USA.}
 \author{R. Kaur}
 \affiliation{Department of Materials Science and Engineering, The Ohio State University, Columbus, OH 43210, USA.}
 \author{M. Eder}
 \affiliation{Department of Physics, The Ohio State University, Columbus, OH 43210, USA.}
  \author{Y. Noor}
 \affiliation{Department of Materials Science and Engineering, The Ohio State University, Columbus, OH 43210, USA.}
 \author{D. P. Broughton}
 \affiliation{Los Alamos National Laboratory, Los Alamos, NM 87545, USA.}
 \author{Md Kazi Rokunuzzaman}
 \affiliation{Department of Materials Science and Engineering, The Ohio State University, Columbus, OH 43210, USA.}
 \author{R. Hollinger}
 \affiliation{Electrical and Computer Engineering Dept, Colorado State University, Fort Collins, CO 80523, USA.}
 \author{A. Blackston}
 \affiliation{Department of Materials Science and Engineering, The Ohio State University, Columbus, OH 43210, USA.}
 \author{J. Strehlow}
 \affiliation{Los Alamos National Laboratory, Los Alamos, NM 87545, USA.}
  \author{A. Baraona}
 \affiliation{Department of Physics, The Ohio State University, Columbus, OH 43210, USA.}
   \author{P. Spingola}
 \affiliation{Department of Physics, The Ohio State University, Columbus, OH 43210, USA.}
  \author{G. Tiscareno}
 \affiliation{Department of Physics, The Ohio State University, Columbus, OH 43210, USA.}
  \author{D. Hanggi}
 \affiliation{Department of Physics, The Ohio State University, Columbus, OH 43210, USA.}
\author{B. Unzicker}
 \affiliation{Department of Physics, The Ohio State University, Columbus, OH 43210, USA.}
 \author{C.-S. Wong}
 \affiliation{Los Alamos National Laboratory, Los Alamos, NM 87545, USA.}
 \author{G. K. Ngirmang}
 \affiliation{National Sciences and Science Education, National Institute of Education, Nanyang Technological University, Singapore 637616, Singapore.}
 \author{F. N. Beg}
 \affiliation{Center for Energy Research, University of California San Diego, La Jolla, CA 92093, USA.}
\author{D. Schumacher}
 \affiliation{Department of Physics, The Ohio State University, Columbus, OH 43210, USA.}
 \author{E. Chowdhury}
 \affiliation{Department of Materials Science and Engineering, The Ohio State University, Columbus, OH 43210, USA.}
 \affiliation{Department of Electrical and Computer Engineering, The Ohio State University, Columbus, OH 43210, USA.}
 \affiliation{Department of Physics, The Ohio State University, Columbus, OH 43210, USA.}

\date{\today}

\begin{abstract}
Relativistic lasers on solid targets generate hot electrons, and other secondary particles. These particles can be used for radiography, cancer therapy, or isochoric heating. A lower density or structured coating on high-Z targets can improve laser-target energy coupling and subsequently enhance overall particle emission. In this work performed at the Scarlet Facility, a 10\textsuperscript{21}~W/cm\textsuperscript{2} intense pulse was incident on front surface coatings on 1~mm thick Ta. These coatings include a 12~$\mu$m plastic coating, a 50~$\mu$m thick foam coating, and a Au nanowire (NW) coating. Post-damage craters are correlated with reflected light on a MACOR screen, illustrating that less absorption in a target is directly tied to smaller craters. Additionally, more absorption in a target also leads to more MeV electrons and X-rays. Bare targets performed the best for electron and MeV X-ray generation, with X-rays of 30~MeV detected, as coatings tested were too thick and thus experienced lower intensities. Due to this larger spot size, foam and NW-coated targets generated the greatest heavy ion acceleration. Particle-in-cell simulations tested on bare and plastic-coated targets illustrate that $\sim$$\mu$m thick plastic coatings perform better than bare Ta. These results underline the importance of density and thickness control of coatings on high-Z materials. In the future, post-damage crater analysis could provide an easy way to benchmark absorption in a sample, and could later be compared against absorption estimates from particle-in-cell simulations.
\end{abstract}

\maketitle

\section{Introduction}
\label{sec:intro}
A high intensity laser (>10\textsuperscript{18}~W/cm\textsuperscript{2}) incident on a solid target may accelerate electrons relativistically \cite{courtois2011high,singh2021bremsstrahlung}, primarily through \textit{j$\times$B} acceleration \cite{wilks1992absorption,malka1996experimental}. These energetic electrons can then seed secondary particles such as: ions\cite{jung2011novel}, protons \cite{vallieres2021enhanced}, positrons \cite{chen2009making}, or MeV X-rays \cite{strehlow2025filter,Bogale_2025}. Ion beams can be used for isochoric heating \cite{fernandez2017laser}, cancer therapy \cite{bulanov2002oncological, fiuza2012laser}, radiography \cite{borghesi2002electric}, and neutron generation \cite{roth2013bright}. Protons have been used succesfully for radiotherapy of specific cancers \cite{verma2016systematic}. Meanwhile, X-rays have been utilized in nuclear physics, medicine, radiation effects, or flash radiography \cite{courtois2009effect,strehlow2024mev}. In particular, X-rays generated from high-Z targets have been used for radiography as an alternative to traditional X-ray sources. Bremsstrahlung X-ray sources driven by short pulse lasers have a spot size of $\sim$100~$\mu$m, tunable angular distribution, ultrashort duration, and can range up to MeV levels \cite{courtois2011high, courtois2013characterisation, strehlow2024mev}. Radiographs taken with a smaller spot size have higher resolution than conventional sources, where spot size is $\sim$mm \cite{courtois2011high}. Additionally, these energetic X-rays may be used to radiograph high density objects such as double-shell ignition targets \cite{courtois2013characterisation}, which cannot be penetrated by lower energy sources.

Traditionally, lasers on bare, solid targets can be plagued by low laser-target energy coupling. The main pulse is typically preceded by a prepulse, which when incident on the target ionizes the surface, thus creating an expanding plasma that follows a decaying exponential. As the main pulse propagates through the plasma it encounters an overdense plasma ($\omega$\textsubscript{p} > $\omega$\textsubscript{l}) from which the laser pulse is reflected. \cite{palaniyappan2012dynamics}. At high intensities it is possible to increase laser-target coupling through relativistic transparency, where relativistic electrons on the sample have increased mass and render a plasma underdense, but it requires high temporal contrast and specific target geometry \cite{palaniyappan2012dynamics,fernandez2017laser}. However, for bare targets with high intensity lasers, the target may only absorb up to 10\% of the laser's energy \cite{PurvisThesis2014,li2022high} due to an overdense plasma being reached more quickly thus limiting absorption to a thin surface layer \cite{wu2021dynamics}.

A coating may be used as the front layer of a high-Z target to thus improve laser-target coupling. A coating of nanowires (NWs) or low density aerogels can be used to increase the coupling to 79 - 88\% \cite{eftekhari2022laser,li2022high}. In these structured targets, more electrons can be accelerated into the high-Z substrate before the plasma optically shutters the peak of the pulse \cite{cao2010enhanced}. A thin, solid-density plastic coating can also be used to increase laser absorption into the target, by creating a homogeneous plasma for the laser to interact with \cite{singh2021bremsstrahlung,strehlow2024mev}. Generally, a solid plastic coating may be more appealing than structured coatings, as structured coatings have more parameters that need to be optimized (such as nanowire length, diameter, density, or a foam thickness and density) \cite{vallieres2021enhanced,li2022high}, and are less costly to produce.

 In this work conducted at the Scarlet Facility, a 815~nm, 50~fs, 5-7~J pulse was incident on 1~mm thick Ta. Plastic, aerogel/foam, and NW coatings were all tested against bare Ta. Peak intensities on target ranged from 0.5-4$\times$10\textsuperscript{21}~W/cm\textsuperscript{2}, with pre-pulse contrast of 10\textsuperscript{-8} at -75 ps and 10\textsuperscript{-6} at -25 ps. Generated electrons, Ta ions, and MeV X-rays were recorded and characterized simultaneously for each individual target type. In addition, qualitative laser-target coupling efficiency can be described by use of a diffuse reflector (MACOR). Post-damage crater measurements are compared against MACOR images and can offer insight into laser-target energy coupling. 
 
 MeV X-rays have not been characterized before at this facility, and both the high contrast and short pulse duration are unusual in these hard X-ray experiments \cite{fernandez2017laser,strehlow2025filter}.  Nanostructured-coatings are rarely used in these high-Z targets, and more insight into their behavior could prove fruitful for future studies \cite{singh2021bremsstrahlung, tavana2023ultra}. High laser contrast is crucial with structured targets, as a low-intensity pre-pulse could generate a critical density surface before the arrival of the peak of the pulse \cite{purvis2013relativistic}. These results will continue to fill in gaps in the collective understanding about relativistic laser-matter interactions and its subsequent emissions.

\section{Experimental setup}
\label{sec:setup}

\begin{figure}[hbt!]
\centering
\fbox{\includegraphics[height=.5\textwidth]{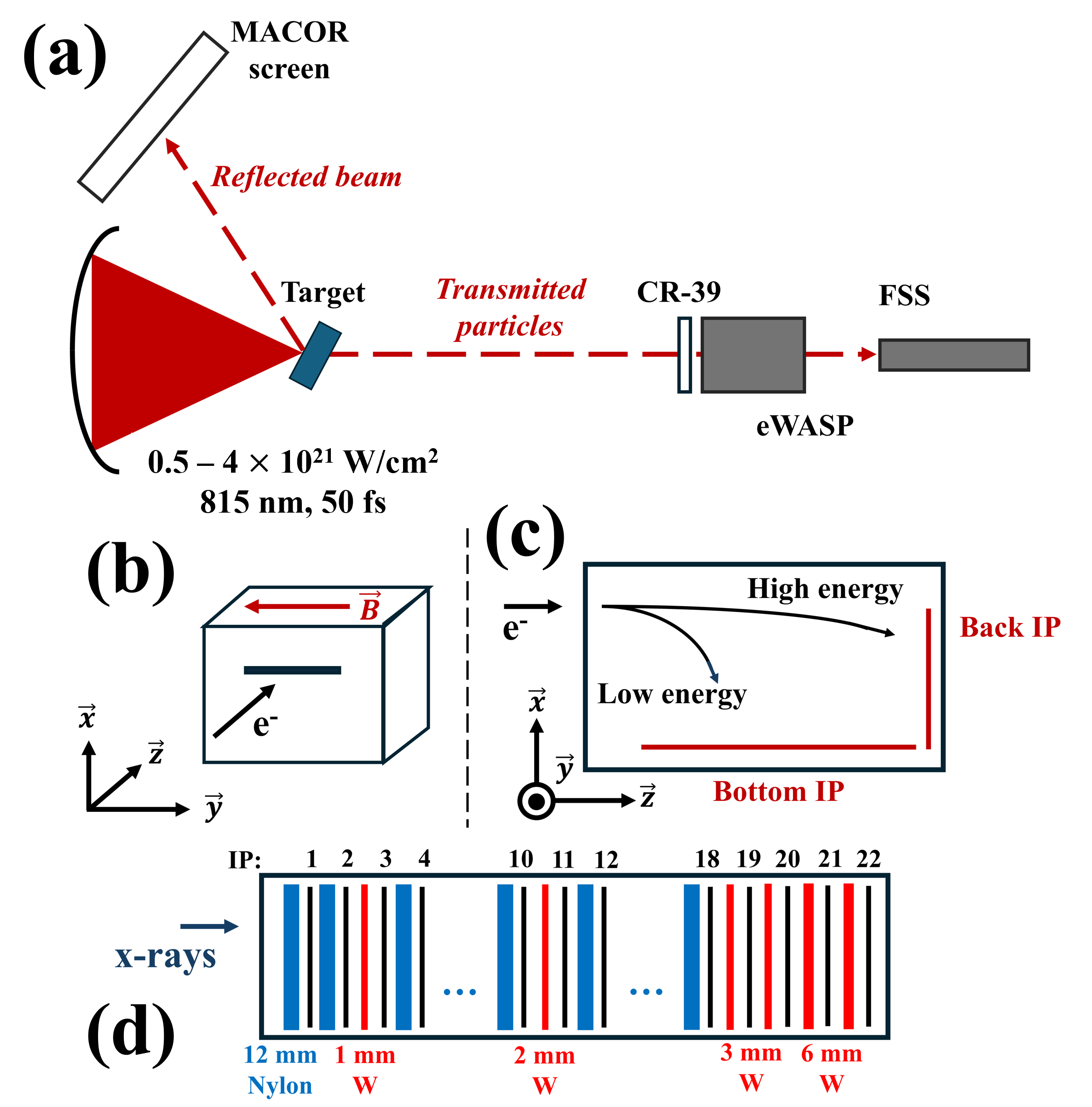}}
\caption{(a) Experimental setup with an angle of incidence of 28\textdegree. The MACOR screen is rotated 50\textdegree, and was placed directly next to the focusing optic. The CR-39 is placed behind the target in the laser axis, with a horizontal slit cut out for the electron wide angle spectrometer (eWASP). The filter stack spectrometer (FSS) is placed outside of the chamber and in the laser axis. (b) Front view of 5.1 $\times$ 5.1~cm CR-39, and its mylar coating thickness. (c) Side view of eWASP, where the electrons enter the slit in +$\protect\vv{z}$, $\protect\vv{B}$ is in -$\protect\vv{y}$. (d) Side view of FSS, with the IP numbers labeled above the FSS. The majority of the filters are nylon-66, while some are tungsten (W).}
\label{fig:Setup}
\end{figure}

\begin{figure}[hbt!]
\centering
\fbox{\includegraphics[height=.3\textwidth]{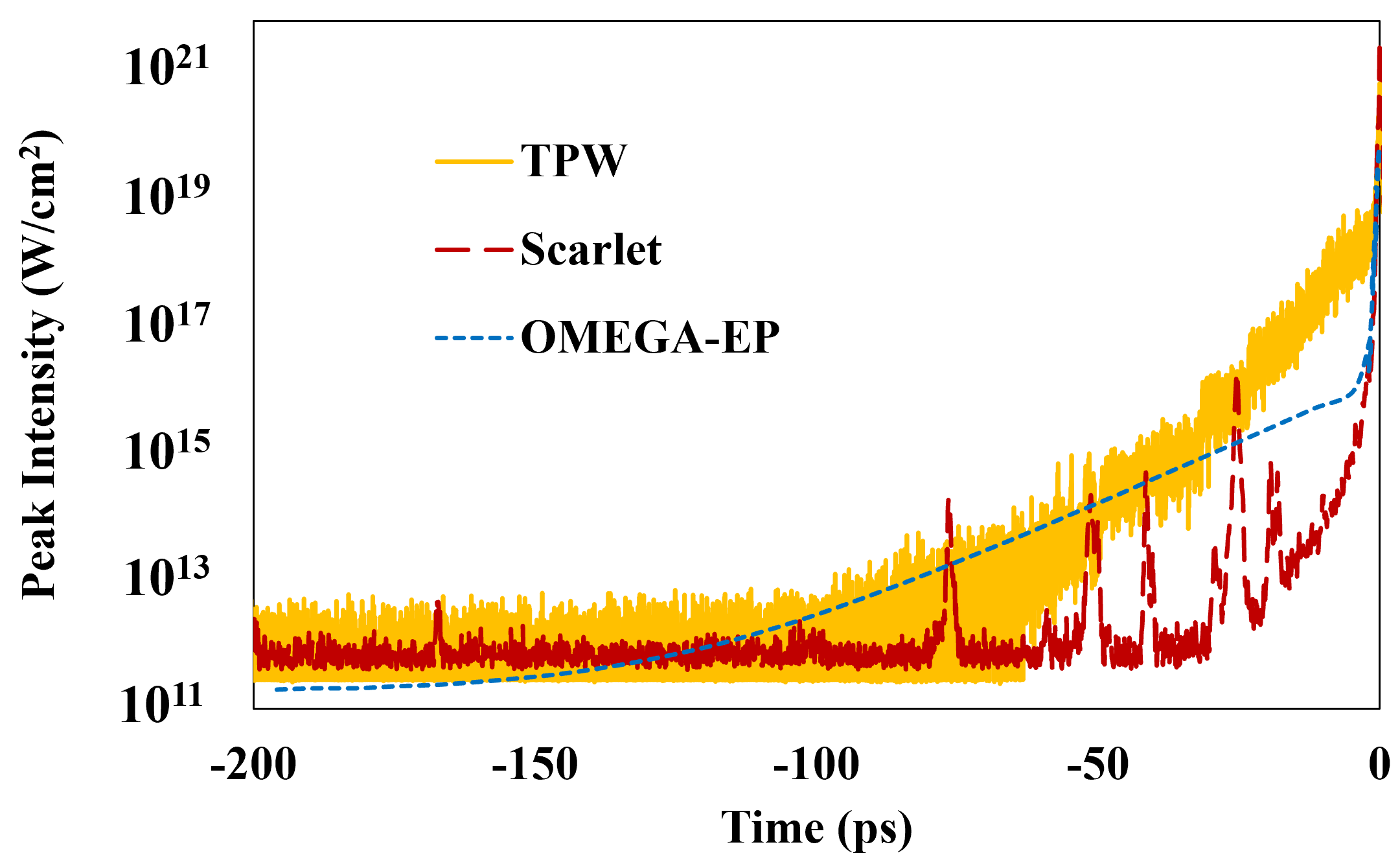}}
\caption{Measured pre-pulse contrast of Scarlet (cross correlation) versus other high power lasers, with the peak of the pulse occurring at 0~ps.}
\label{fig:Contrast}
\end{figure}

 At the Scarlet Facility (LasernetUS experiment K-10019), a \textit{p}-polarized, 815~nm, 50~fs, 5-7~J pulse was incident at 28\textdegree~on 1~mm thick bare and coated Ta targets. With an f/2 OAP, the focal spot had a \textit{1/e\textsuperscript{2}} diameter of 4.2~$\mu$m. Peak intensities on target were from 0.5-4$\times$10\textsuperscript{21}~W/cm\textsuperscript{2}, with pre-pulse contrast of 10\textsuperscript{-6} at -25 ps. The experimental setup is shown in Fig. \ref{fig:Setup}(a), while the contrast of the laser is featured in Fig. \ref{fig:Contrast}. All targets tested were on a 1~mm thick Ta substrate, with the laser focused on the Ta substrate. Coatings tested were: 12~$\mu$m thick photoresist (AZ nLOF2070) with 1.07~g/cc (spin-coated at 500~RPM, from Ohio State University (OSU)), 50~$\mu$m thick aerogel (GA-CH) with 15.4 mg/cc (from General Atomics), Au nanowires (NWs) aligned vertically with a diameter of 0.51 $\pm$ 0.06~$\mu$m and a length of 4.39 $\pm$ 0.50~$\mu$m (from Colorado State University(CSU)), alongside bare Ta. SEM images of the foam and NW-coated targets are featured in the supplementary materials, Fig. s1. Each different target type was mounted on a rotating wheel that held six targets, so all of the generated electron spectra, X-ray spectra, and ion histograms were cumulative for all six shots. By using this sample holder and a mechanized stage, it was possible to shoot all six targets in one vacuum pump cycle, and the image plates and CR-39 were collected after each set of shots. Target alignment was performed on the front side of the target, using a microscope objective that was moved out with a mechanized stage during system shots.

Behind the target in the laser axis is mylar-coated CR-39. The mylar is laid in a stepwise fashion over the front, ranging in thickness from 0 - 12~$\mu$m which can then discern Ta ions from 0 - 60~MeV (Fig. \ref{fig:Setup}(b)). Protons are visible on etched CR-39, but various energy levels cannot be discerned. The CR-39 has a horizontal slit cut out, 2.5$\times$26~mm, allowing electrons to pass through to the electron wide angle spectrometer (eWASP). The slit was cut out of the CR-39, but mylar was intact over the slit. After being exposed, the mylar was removed and the CR-39 is put in 6~M NaOH at 95\textdegree~C for 90~minutes. Afterwards, the CR-39 is put in 100\% glacial acetic acid for 90~seconds. Imaging was carried out under an optical microscope at 10x magnification. Ion counts were measured at high resolution for the entire sample from a stitched-together image (via homemade machine-learning Python script).

The eWASP is home-built from 316 stainless steel while the sides are mild steel. The front plate is 2.5~cm thick, while the slit is 0.5~mm in width, and 25~mm across. Modeling of the electron trajectories was carried out in Python, and a MPT-132-7s hall probe was used to measure the magnetic field inside the eWASP. The \textit{B} field deflects the electrons downward (-$\protect\vv{x}$ in Fig. \ref{fig:Setup}(c)) where their final position is determined by their initial energy. The distance between the magnets is 37~cm, leading to a peak magnetic field in the center of the eWASP of 0.35~T. Image plates \cite{izumi2006application} (IPs, BAS-IP SR) were placed at the bottom of the detector, and in the back, while no steel back plate was used (Fig. \ref{fig:Setup}(C)). The eWASP can detect electrons from 1 - 50~MeV.  

The filter stack spectrometer (FSS) is behind the eWASP and outside of the target chamber. The FSS is machined out of tungsten, and carries a stack of  alternating filters and image plates. X-rays are attenuated by the filters and deposit their energy on the IPs, building a diagnostic signal. A new design of the filter stack recipe implemented in this work, prioritizes the Compton electrons produced in the high-Z materials to produce more discerning signals that may be unfolded with greater accuracy (see Fig. \ref{fig:Setup}(d)) \cite{Wong_2024}.

A MACOR (glass-ceramic) screen, which functioned as a diffuse reflector, was used to capture the reflected light off of the target. The screen is 28 $\times$ 28~cm, and turned to 50\textdegree, 36.5~cm away from the target. The MACOR screen captures the majority of the back-reflected light, and thus can be used to make qualitative observations about laser-target coupling between various targets. A camera outside of the chamber triggered at laser time, with a 10~ms exposure images this screen with a zoom lens. 

\section{Experimental Results}
\label{sec:Results}
\subsection{Laser-target coupling}
\label{sec:MACOR}

\begin{figure}[hbt!]
\centering
\fbox{\includegraphics[height=.35\textwidth]{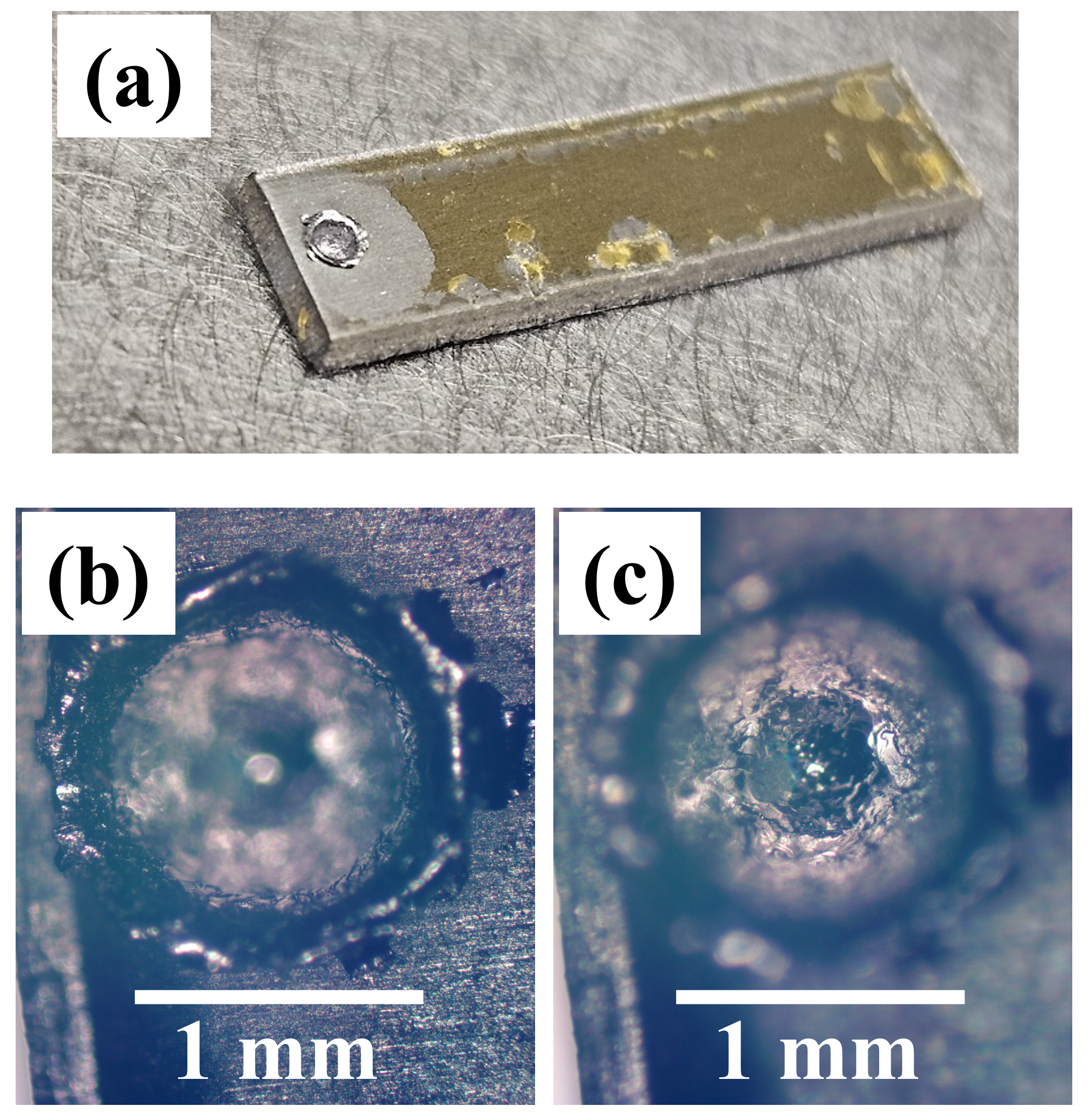}}
\caption{(a) Example crater on plastic-coated Ta. Ta strip is 5 $\times$ 20 $\times$ 1 mm. Crater on bare Ta target with the (b) surface in focus, (c) bottom of crater in focus. Both crater depth and diameter varied with the target tested. Note that the focal spot had a \textit{1/e\textsuperscript{2}} diameter of 4.2~$\mu$m}
\label{fig:Post}
\end{figure}

Energy and peak intensity incident on each target wheel is detailed in Table \ref{tab:Parameters}. Each type of target was shot six times in a single vacuum pump cycle. An example crater from a shot on bare Ta is featured in Fig. \ref{fig:Post}. 

\begin{table}[H]
\caption{\label{tab:Parameters}Average Laser Parameters}
\begin{ruledtabular}
\begin{tabular}{cccc}

 Target & Energy (J) &  Pulse duration (fs) & I\textsubscript{p} ($\times$10\textsuperscript{21} W/cm\textsuperscript{2}) \\
 \hline
  Bare   & 5.43 \textpm~0.36 & 50 &3.45 \textpm~0.23\\
  Plastic   & 5.97 \textpm~0.45 & 60 & 3.16 \textpm~0.24\\
  Foam   & 6.70 \textpm~0.34 & 45 & 0.51 \textpm~0.03\textsuperscript{a}\\
  NWs   & 6.29 \textpm~0.37 & 47 & 3.85 \textpm~0.23\textsuperscript{b}\\
\end{tabular}
\end{ruledtabular}
\footnotesize\textsuperscript{a}\,Lower intensity at surface of foam due to focusing on substrate.

\footnotesize\textsuperscript{b}\,Intensity on surface of NWs is lower but cannot be calculated due to NW geometry.
\end{table}

\begin{figure}[hbt!]
\centering
\fbox{\includegraphics[height=.24\textwidth]{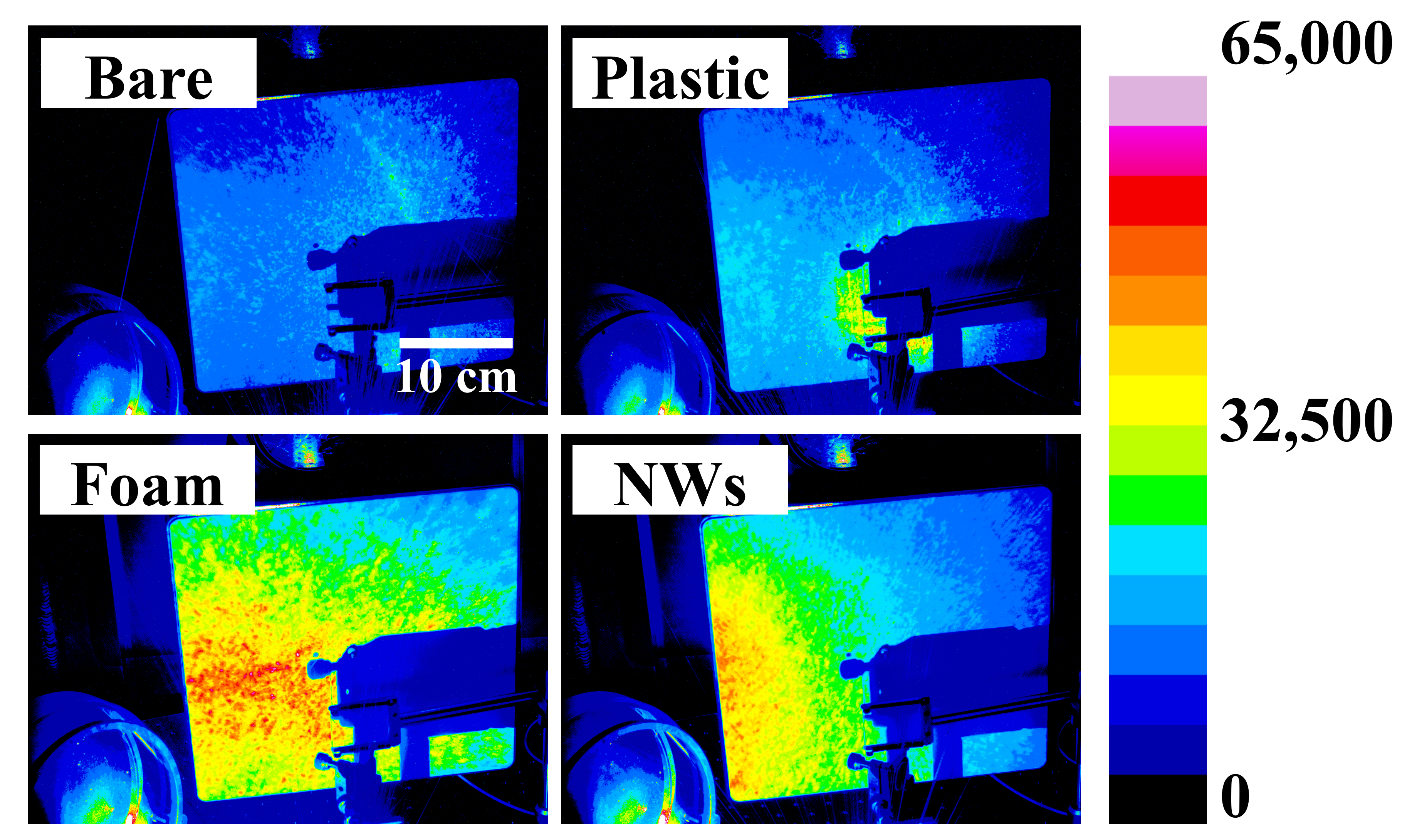}}
\caption{Representative MACOR images for all samples tested. The f/2 focusing optic is in the bottom left of each image. The microscope objective used for front-side imaging is casting a shadow on the MACOR screen, and partially obscuring the front CR-39. MACOR images of NWs were the most variable in intensity.}
\label{fig:MACOR}
\end{figure}

Images obtained from the MACOR screen are featured in Fig. \ref{fig:MACOR}. Only the NW targets generated variable MACOR images, with half of the shots saturating the screen (camera had ND=1.9). Other target types tested produced similar back-reflected scatter on each shot. A blackfly camera (BFS-U3-120S4M-CS) with a zoom lens imaged the MACOR screen, and was located outside of the chamber, 154~cm from the screen. The camera was angled at 45\textdegree~downwards. Due to saturation of the screen, and its location in the chamber, quantitative measurements of absorbed energy in the target cannot be calculated; however, qualitatively, bare and plastic-coated targets appeared to have absorbed more laser energy than the nanostructured targets. 

Post-damage crater morphology reveals that crater size varies with target type, and crucially agrees with images of the MACOR screen. Average crater diameter, depth, and estimated volume are detailed in Table. \ref{tab:Crater}. The volume is obtained with the ellipsoid volume equation: $V = (4/3)\pi abc$, where \textit{a,b,c} are the three radii of the crater. This estimated crater volume is half of the calculated ellipsoid value, since the crater is a hemisphere. Crater depth is measured with a Mitutoyo 543-793-10 digimatic indicator. The MACOR images with the lowest signal features the greatest absorption by the target, also had the largest crater diameters. Additionally, the two brightest MACOR images from the NW-coated targets had the lowest crater diameters measured at 0.56\textpm~0.03~mm. This relationship between MACOR signal and crater size is further detailed in the supplementary materials, Fig. s2.

At high intensities, a phase explosion may occur in which an overheated surface region of the melt layer decomposes into vapor and liquid droplets \cite{wu2014microscopic}. In this regime the surface layer in the highest-fluence region becomes thermodynamically unstable and releases a mixture of vapor and droplets, while material removal outside the central, highest-fluence region is often governed by photomechanical processes (spallation) driven by relaxation of laser-induced stresses \cite{wu2014microscopic,shugaev2016fundamentals}. Because ultrashort pulses deposit energy into a very shallow absorption region set by the optical absorption depth and the ballistic transport of excited electrons, the initial vapor/plasma formation of phase explosion is confined to a thin surface layer (nm–100s nm, depending on material and fluence), whereas deeper material is typically melted and then ejected as droplets or spalled chunks rather than being converted to vapor \cite{wu2014microscopic, zhigilei2000microscopic}. Material removal deeper into the crater can also involve laser-driven hole boring. Hole boring arises when the radiation pressure (ponderomotive force) of the laser pushes the critical-density surface inward, driving a collisionless shock into the plasma and accelerating ions \cite{weng2012ultra}. In this process, the laser-driven plasma layer acts like a piston pushing into the material. At sufficiently high intensities, relativistic transparency can modify the hole-boring dynamics by allowing the laser to propagate deeper into the overdense plasma. High-intensity hole boring in the incomplete hole-boring regime (where the piston velocity is insufficient to expel the entire target thickness before the laser pulse ends) is associated with increased fast-ion cutoff energies \cite{weng2012ultra}.

\begin{table}[H]
\caption{\label{tab:Crater}Average Crater Dimensions versus Target Type}
\begin{ruledtabular}
\begin{tabular}{ccccc}

 Target & Diameter (mm) &  Depth (mm) & Volume (mm\textsuperscript{3}) & Energy Abs. (J) \\
 \hline
  Bare   & 1.46 \textpm~0.08 & 0.48 \textpm~0.05 &0.51 \textpm~0.04 & 5.03 \textpm~0.38\\
  Plastic   & 1.52 \textpm~0.09 & 0.56 \textpm~0.09 & 0.64 \textpm~0.18 & 6.36 \textpm~1.80\\
  Foam   & 1.24 \textpm~0.07 & 0.42 \textpm~0.02 &0.32 \textpm~0.04 & 3.22 \textpm~0.37\\
  NWs   & 0.95 \textpm~0.34 & 0.34 \textpm~0.13 & 0.23 \textpm~0.20 & 2.30 \textpm~1.95\\

\end{tabular}
\end{ruledtabular}
\end{table}

Energy absorbed in Table \ref{tab:Crater} can be obtained by using the specific heat of tantalum, 25.36 J/mol K at 298 K, and the heat of fusion, 31.4 kJ/mol \cite{chase1998nist}, to approximately estimate the minimum required energy required to melt this volume of material. These estimates neglect 1) the vaporized surface layer of the material (10s to 100s of nm \cite{wu2014microscopic}) in a phase explosion, 2) mechanically ejected fragments produced by subsurface void nucleation and growth during the relaxation of high thermoelastic stresses under stress-confinement conditions \cite{zhigilei2000microscopic}, and 3) Coulomb explosion, where a positively charged surface left by the laser ionization and the neutralization by free electrons from the bulk \cite{shugaev2016fundamentals}. If we had assumed thermal equilibrium, with the material boiling off into vapor, the energy required to vaporize this amount of material would require 35.25 ± 2.68 J for the bare targets. This is evidence that the material ejected or removed from the crater is not due to a thermal process.

\subsection{Electrons}
\label{sec:electrons}
\begin{figure}[hbt!]
\centering
\fbox{\includegraphics[height=.3\textwidth]{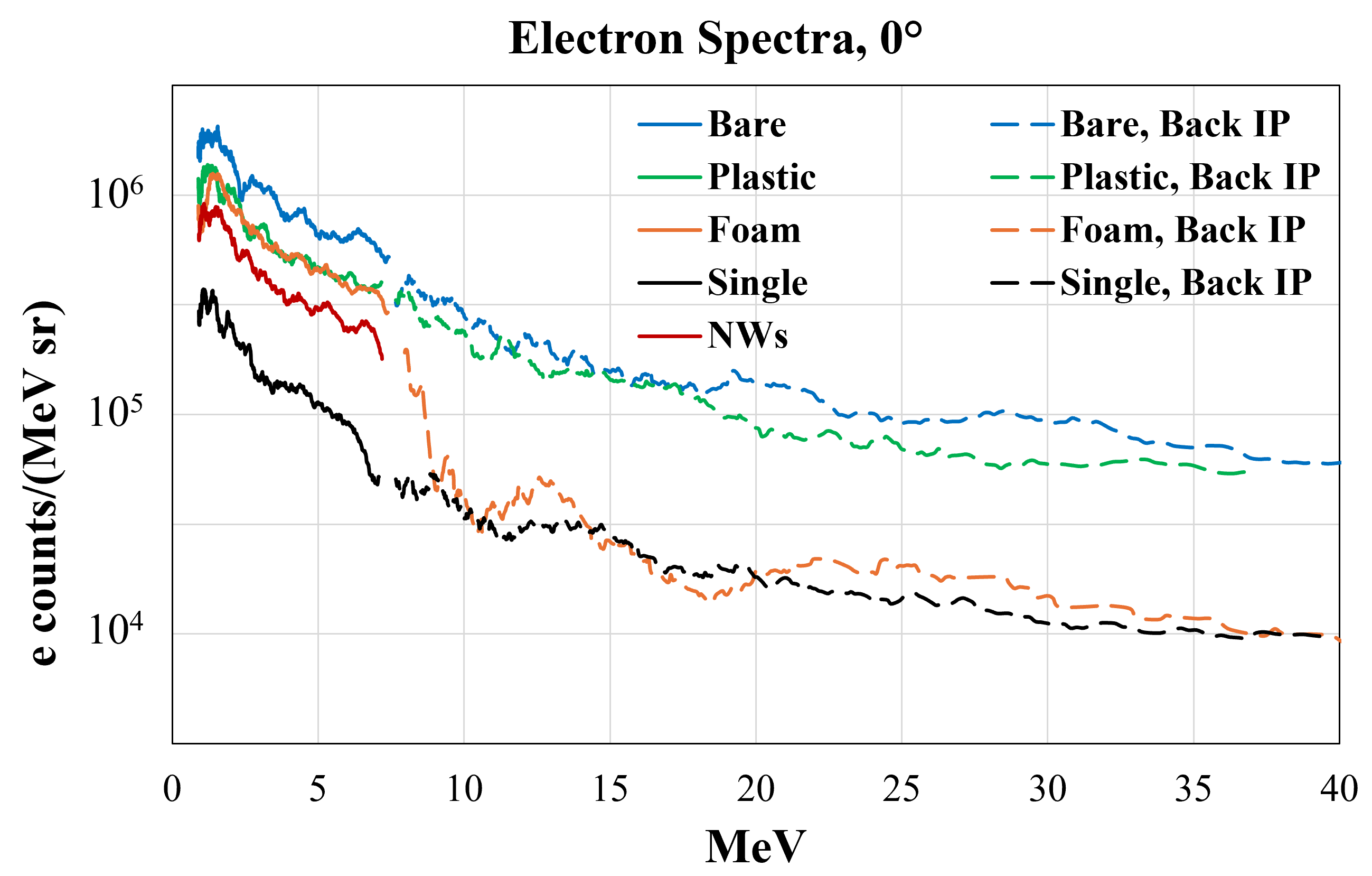}}
\caption{Electron spectra taken at 0\textdegree, directly along the laser axis. The dashed line is the same target wheel, but the back IP instead of the bottom. A single pump shot on bare Ta is featured. The back IP for the foam target may have been loaded incorrectly into the detector. The back IP for the NW target was not loaded correctly and is not shown.}
\label{fig:eSpectra}
\end{figure}

Featured is the electron spectra obtained from the eWASP (Fig. \ref{fig:eSpectra}) as well as the average electron temperature obtained from fitting to an exponential distribution: $f \approx exp(-E/T)$ (Table \ref{tab:eSpectra}) where \textit{E} is the electron energy and \textit{T} is the derived temperature. The average electron temperatures were obtained using the bottom and back IPs of the eWASP, which captured electrons up to $\sim$50~MeV. Due to the distance from the target to the detector of 33.8~cm, only \textpm1.5\textdegree~could be observed. The eWASP could not be closer to the target due to routine focal spot imaging.

Total number of electrons incident on the eWASP was highest for the bare targets, as the IP had to be scanned 14~times to remove saturation. The plastic-coated targets had the next highest electron count, and the IP had to be scanned 9~times. Finally the foam and NW targets both needed to be scanned 6~times, and produced the fewest electrons on the eWASP. The spectra in Fig. \ref{fig:eSpectra} are scaled according to the number of times the particular IP scanned as the IP lost $\sim$10\% signal each time it was scanned.

\begin{table*}
\caption{\label{tab:eSpectra}Fitted Electron Temperatures}
\begin{ruledtabular}
\begin{tabular}{cccccc}

 Angle (\textdegree) & Bare, T(MeV) & Plastic, T(MeV)& Foam, T(MeV) & NWs, T(MeV) & Single Bare\footnote{This is from a single bare target, instead of six.}, T(MeV)\\
 \hline
  -1.50   & 5.30 \textpm~0.21 &5.12 \textpm~0.20&5.19 \textpm~0.17&4.37 \textpm~0.17&4.80 \textpm~0.19\\
 -0.75   & 5.07 \textpm~0.21 &4.24 \textpm~0.15&4.74 \textpm~0.15&4.06 \textpm~0.16&3.62 \textpm~0.12\\
 0   & 4.81 \textpm~0.22 &3.93 \textpm~0.14&4.72 \textpm~0.17&3.48 \textpm~0.14&3.62 \textpm~0.12\\
 0.75   & 4.87 \textpm~0.20 &4.41 \textpm~0.15&4.75 \textpm~0.16&3.68 \textpm~0.13&3.55 \textpm~0.11\\
 1.50   & 5.00 \textpm~0.20 &4.78 \textpm~0.21&5.89 \textpm~0.23&4.59 \textpm~0.21&4.85 \textpm~0.21\\
 \hline
 Average   & 5.01 \textpm~0.21 &4.50 \textpm~0.17&5.06 \textpm~0.18&4.03 \textpm~0.16&   4.09 \textpm~0.15\\
 Average, Back IP (T\textsubscript{hot})   & 17.77 \textpm~0.94 &15.22 \textpm~0.77&10.63\footnote{The specific IP may have been loaded incorrectly into the eWASP} \textpm~0.99& - &   17.63 \textpm~1.30\\

\end{tabular}
\end{ruledtabular}
\end{table*}

\subsection{MeV X-rays}
\label{sec:FSS}
While the electrons measured by the eWASP reveals information regarding the laser-plasma acceleration processes, the electrons accelerated near the front surface of the target are depleted as they traverse the 1 mm Ta target. Additional information can be recovered by assessing the secondary X-rays produced via bremsstrahlung processes. The challenge occurs with the recovery of the spectrum. Conventional techniques for keV X-rays such as crystal-based spectrometers and X-ray streak cameras, become ineffective at MeV levels. Diagnostics that have effectiveness in this MeV range, such as Compton-based spectrometers and single-photon counting cameras, require fluxes that are either too high or too low, respectively, for the flux space occupied by laser-driven sources. To address these issues, a diagnostic routine has been developed utilizing the FSS and a perturbative minimization unfolding algorithm to recover the spectrum \cite{Wong_2024}. 

Typical attenuation-based X-ray diagnostics rely on a series of filters and image plates to attenuate photons of different energies in different amounts from which a response matrix can be built using particle transport codes, such as Monte Carlo N-Particle Transport (MCNP) code~\cite{TechReport_2017_LANL}. The raw photostimulated luminescence (PSL) signal in conjunction with the characteristic response matrix may then be used to unfold the spectrum by solving the inverse of the following \textit{\textbf{$\sum_{}^{} S_{i}RM_{ij} = PSL_{j}$}}, where \textit{S} is the vector containing counts in the X-ray spectrum, and \textit{RM} is the response matrix for the FSS \cite{Wong_2024}. Although this is effective in the keV range, in the MeV range the mass-attenuation coefficients flatten out, making it difficult to discriminate between photons of different energies, resulting in an ill-conditioned matrix. A more extensive discussion is detailed by Wong et al. \cite{Wong_2024}.

The work presented here builds on previous studies involving the Filter Stack Spectrometer (FSS). Typically, to differentially filter the X-ray signal, low-Z and high-Z materials are placed towards the front and back of the stack, respectively. A byproduct of MeV photon interactions with these high-Z filters is the generation of Compton electrons. While these electrons are typically accounted for in MCNP calculations, their potential has not been fully utilized. Recent work has shown that a filter stack specifically designed to generate electrons consistently throughout the stack can result in a response matrix that is better conditioned and subsequent unfolding of a higher-fidelity X-ray spectrum\cite{Bogale_2025}. The response matrix from this Compton-based X-ray spectrometer is shown in the supplementary materials (Fig. s3). 

\begin{figure}[hbt!]
\centering
\includegraphics[width=0.92\columnwidth]{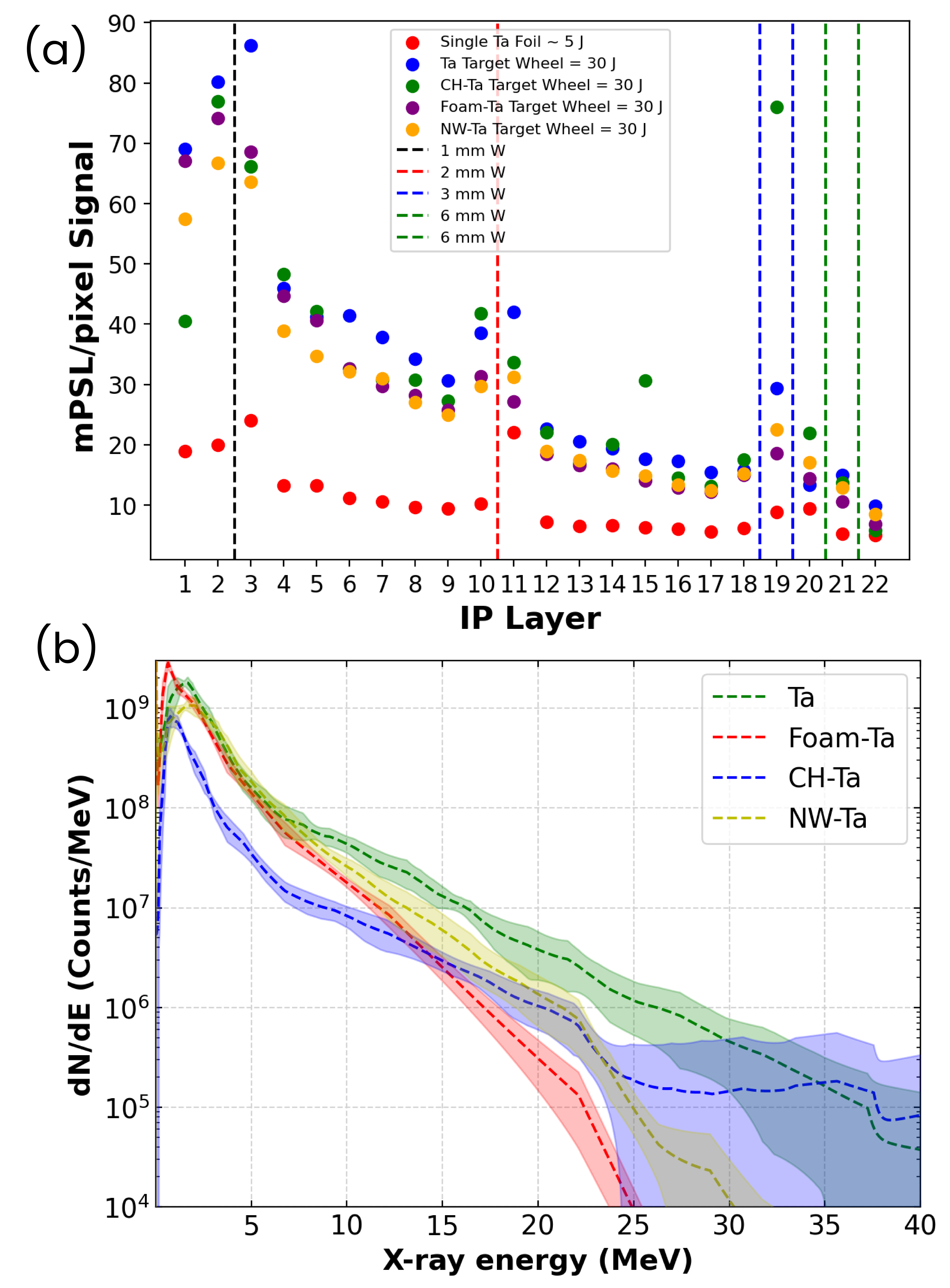}
\caption{(a) The raw PSL per pixel signal is shown for the different target types. The vertical dashed lines represent the W filters, which act as intensifiers via the production of Compton electrons. The enhanced signal from the back and forward scatter electrons is apparent. (b) Final X-ray spectrum. The unfolding routine uses the PSL data to recover the X-ray spectrum.}
\label{fig:unfold}
\end{figure}

The raw PSL signal and MeV X-ray spectra from the different target types via our unfolding routine can be found in Fig. \ref{fig:unfold}(a) and Fig. \ref{fig:unfold}(b), respectively. Dose per total shot energy for bare, plastic-coated, foam-coated, and NW-coated are 4.2$\times$10\textsuperscript{-6}~rad~@1m/J sr, 3.3$\times$10\textsuperscript{-6}~rad~@1m/J sr, 3.0$\times$10\textsuperscript{-6}~rad~@1m/J sr, and 3.0$\times$10\textsuperscript{-6}~rad~@1m/J sr respectively. 

\subsection{Ta ions}
\label{sec:Ions}

\begin{figure}[hbt!]
\centering
\fbox{\includegraphics[height=.48\textwidth]{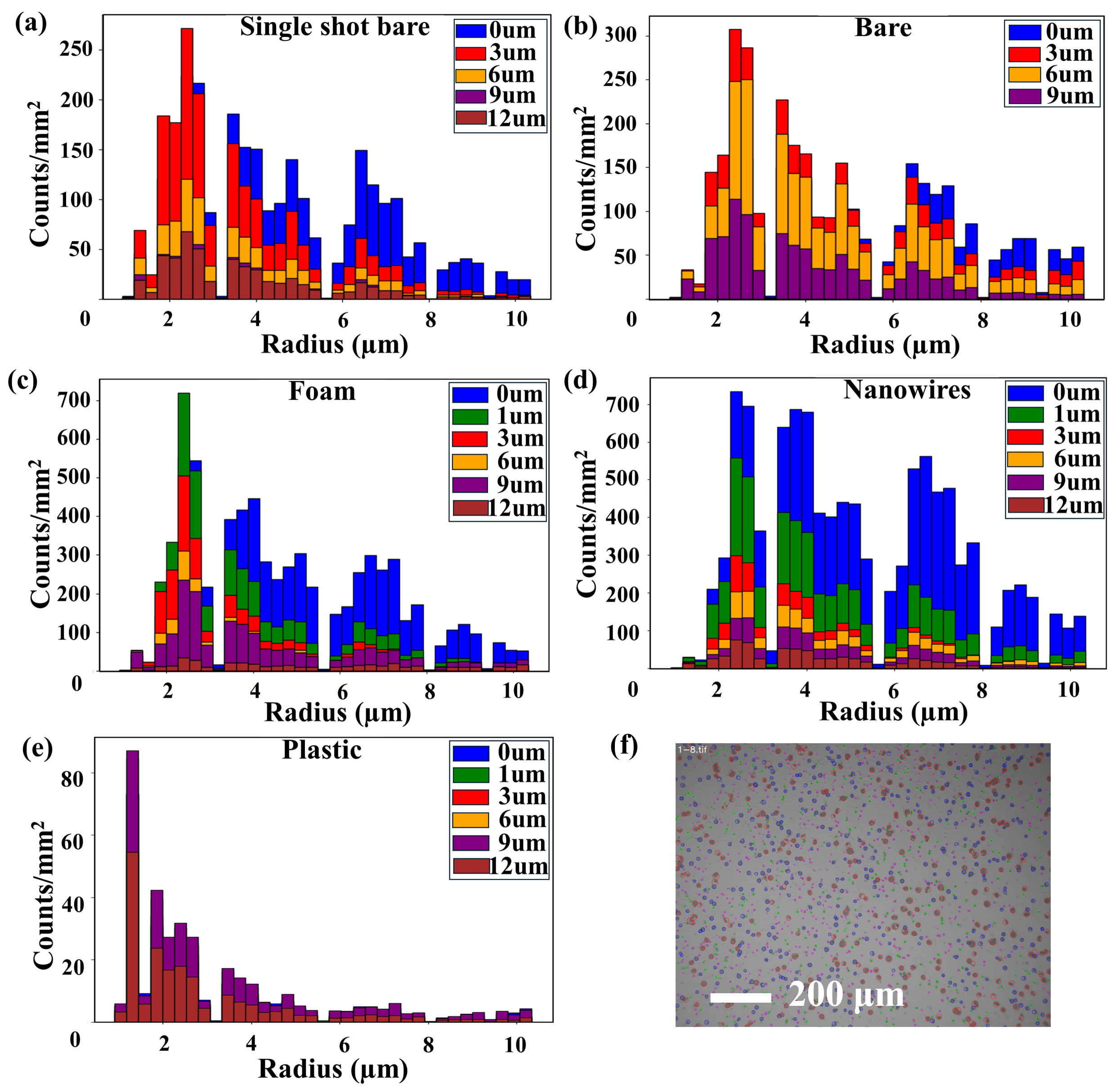}}
\caption{(a)-(e) Experimental counts of Ta ions, with each pit radius measured by a machine-learning technique. Each CR-39 was exposed for six shots on each target type (except the single shot bare), with different thicknesses of mylar to differentiate Ta ions based on energy. (f) Example microscope image of rear CR-39 with the pits counted.}
\label{fig:Histogram}
\end{figure}

The main results from the attenuated CR-39 is featured in Fig. \ref{fig:Histogram} and Table \ref{tab:ionSpectra}. CR-39 was positioned in front of the MACOR screen in the reflected direction off the target, and behind the targets; however, most of the front CR-39 was obstructed by a microscope objective used for sample alignment and the only exposed part did not have any mylar shielding. These CR-39 were saturated, but sections were counted by hand to provide a low-end estimate for total ion counts. The front CR-39 for the single shot on bare Ta was also saturated. Front CR-39 was located 33~cm from the target, at an angle of $\sim$48\textdegree, while the back CR-39 was 31~cm from the target.

The mylar thicknesses were chosen to be able to discern Ta ion energies via SRIM (Stopping and Range of Ions in Matter), with 12~$\mu$m mylar allowing ions > 45~MeV through. The energy cutoff for each thickness of mylar is featured in Table \ref{tab:MylarThickness}. Proton energies could not be differentiated because of this inadequate shielding, but could theoretically be counted under optical microscope. Counting of protons would require 50x magnification instead of the 10x used for the ions. Unfortunately, this range of mylar thickness also allows C ions to be differentiated between 0.35-12 MeV. As the front of the target was coated, the front CR-39 likely detected more C ions, these could have been differentiated by multi-step etching. 

\begin{table}[H]
\caption{\label{tab:MylarThickness}Mylar attenuation of Ta ions from CR-39 positioned behind the target}
\begin{ruledtabular}
\begin{tabular}{cc}

 Mylar thickness ($\mu$m) & Allowed energies (MeV) \\
 \hline
  0  & > 0 MeV \\
  1   & > 2.5 MeV \\
  3   & > 7.0 MeV \\
  6   & > 16.0 MeV  \\
  9   & > 27.5 MeV \\
  12   & > 45.0 MeV \\

\end{tabular}
\end{ruledtabular}
\end{table}

\begin{table*}
\caption{\label{tab:ionSpectra}Ion counts/sr J [$\times$10\textsuperscript{7}]}
\begin{ruledtabular}
\begin{tabular}{cc|cccccc}

 Target & Front 0~$\mu$m &Back 0~$\mu$m & 1~$\mu$m & 3~$\mu$m & 6~$\mu$m & 9~$\mu$m & 12~$\mu$m\\
 \hline
  Single Shot Bare   & - & 4.3 & - & 3.0 & 1.6 & 0.80 & 0.80\\
 Bare   & > 10.3$\times$10\textsuperscript{3} & 4.6 & - & 5.2 & 4.2 & 1.8 & -\\
 Plastic   & > 4.2$\times$10\textsuperscript{3} & 0.56 & 0.47 & 0.42 & 0.44 & 0.59 & 0.32\\
 Foam   & > 2.3$\times$10\textsuperscript{3} & 9.5 & 6.3 & 4.1 & 2.7 & 2.6 & 0.58\\
 NWs   & > 4.7$\times$10\textsuperscript{3} & 17 & 8.4 & 4.2 & 3.3 & 2.1 & 1.1 \\

\end{tabular}
\end{ruledtabular}
\end{table*}

Foam and NW-coated targets exhibited an even spread of ions across the 5.1 $\times$ 5.1~cm CR-39, while bare and plastic-coated targets had clear regions of heightened ion density. The reported ion counts in Table \ref{tab:ionSpectra} are all the pits for the entire back CR-39, while three representative microscopy images at 50x were used to estimate total counts for the front CR-39 of each target tested. 

\section{Particle-in-cell simulation}
\label{sec:PIC}

We performed two-dimensional (2D) particle-in-cell (PIC) simulations using version 4.17.10 of the EPOCH code~\cite{Arber2015}, modeling both bare Ta target and Ta coated with a plastic layer. 
The schematic setup for both simulations is shown in the supplementary materials, where the pulse is injected at 28$^\circ$ with a central wavelength of $\lambda$ = 0.8~$\mu$m, temporal sin$^2$ FWHM $\tau$ = 50~fs, \textit{1/e\textsuperscript{2}} radius of $\Omega$\textsubscript{0} = 3~$\mu$m, and a peak intensity of 3~x~10\textsuperscript{21}~W/cm\textsuperscript{2}.
A 10~$\mu$m thick bulk Ta is used to represent the bare Ta case, while a 1~$\mu$m thick plastic coating consisting of carbon (C) and hydrogen (H) is added before Ta to represent the plastic-coated Ta. The pulse is on for a 2$\tau$ window from t = 0~fs. Due to computational constraints, we are unable to model the full 1~mm thick tantalum (Ta) target used in the experiment. Additionally, the thickness of the plastic layer is set to 1~$\mu$m to ensure that it exceeds the laser wavelength $\lambda$, and therefore properly represents the interaction with the laser field.

The material number densities and band gaps used in the simulations are summarized in Table~\ref{tab:mat_prop}. For simplicity, each plastic molecule is assumed to consist of two carbon atoms and four hydrogen atoms, although the actual composition of AZ nLOF 2070 may also include nitrogen (N) and oxygen (O). Tantalum (Ta), which has 73 electrons, cannot be fully ionized within the computational constraints of the simulation. Instead, only the first four ionization states (i.e. Z=4) are included to capture the representative physics and key phenomena. Similarly, for the plastic material, each constituent atom is assumed to ionize up to one electron to simplify the modeling of plasma formation.

These simplifications may underestimate the total electron density and overestimate the average electron energy. However, they preserve the essential plasma dynamics while ensuring computational feasibility. To complement this approach, an additional simulation is performed for bare Ta with the maximum charge state extended to Ta$^{21+}$, corresponding to ionization energies up to about 600~eV, to assess the effects of higher ionization on plasma behavior.

\begin{table}[htb]
\centering
\caption{Material Properties.~\cite{NIST_ASD,arblaster2018crystallographic,lide1995crc}}\label{tab:mat_prop}
\begin{ruledtabular}
\begin{tabular} {ccc}
Material & Ionization Energy (eV) & Number Density (cm$^{-3}$) \\
\hline
Ta &7.55, 16.17, 23.11, 35.03 & 5.55e+22 \\
H  &13.6 & 8.24e+22 \\
C  &13.6 & 4.12e+22 \\
\end{tabular}
\end{ruledtabular}
\end{table}

The simulations for Ta with four ionization states use a spatial resolution of $12~\mathrm{nm}$ in both directions, while the simulation for Ta with 21 ionization states uses resolutions of $7~\mathrm{nm}$ in the \textit{x} direction and $10~\mathrm{nm}$ in the \textit{y} direction. All simulations are initialized with 100 neutral macroparticles per cell for each species at a temperature of 298~K. For each simulation, we use the default time step in 2D EPOCH of 0.95 times the Courant–Friedrichs–Lewy (CFL) limit~\cite{CFL_Paper,CFL_Paper_English,Arber2015}.


\begin{figure}[hbt!]
\centering
\fbox{\includegraphics[height=.37\textwidth]{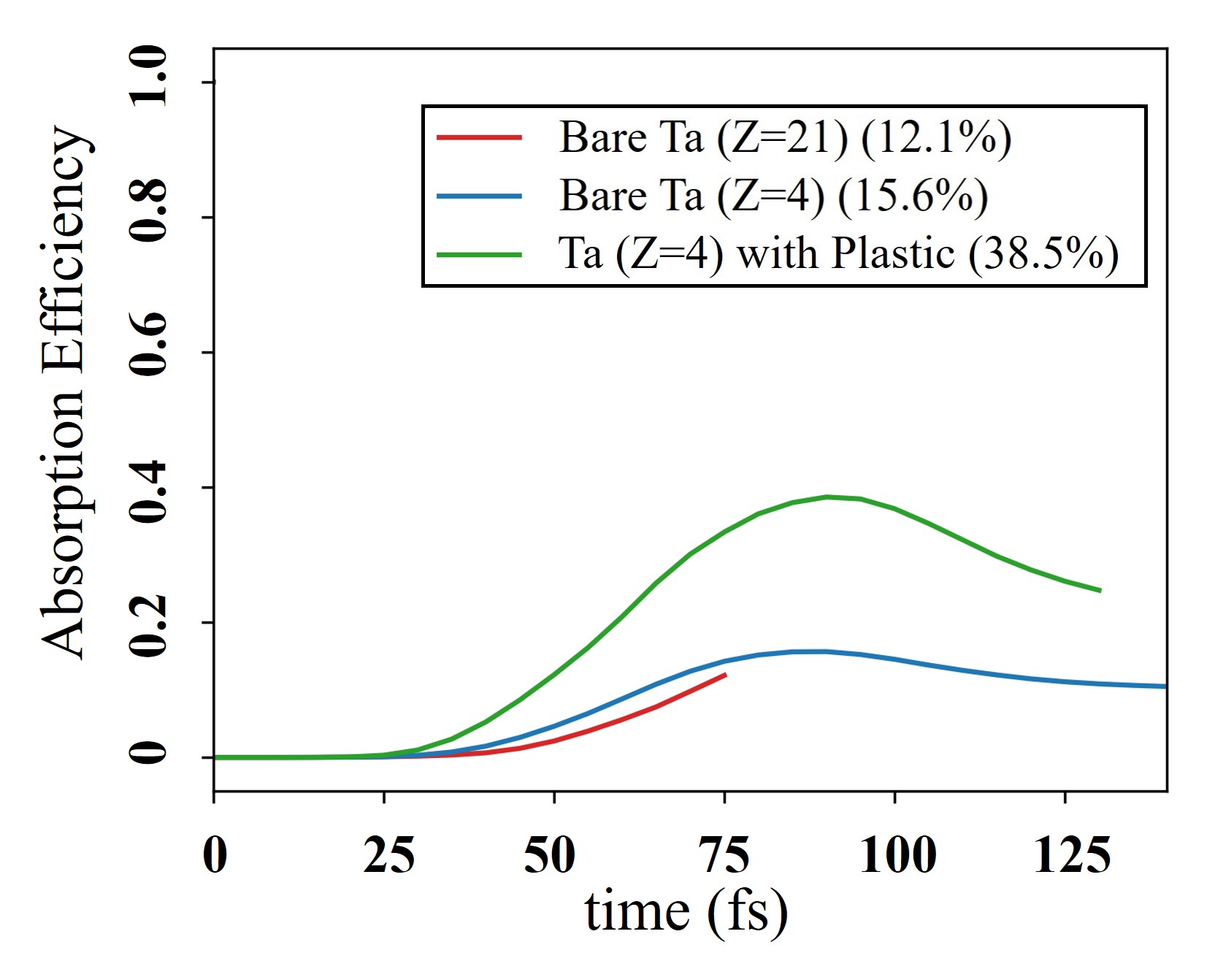}}
\caption{Absorption efficiency for bare Ta(Z=21), bare Ta(Z=4), and Ta(Z=4) with plastic coating, with the maximum efficiency 12.1\%, 15.6\%, 38.5\%, respectively. The bare Ta has an absorption peak slightly earlier than plastic-coated.}
\label{fig:Sim_Absorption}
\end{figure}

The absorption efficiency for all particle species in the three cases is shown in Fig.~\ref{fig:Sim_Absorption}. For bare Ta with 21 ionization states, the extremely large number of electron species significantly increases the computational cost, requiring nearly 35~hours of computation for every 5~fs simulated. This makes it impractical to complete the entire simulation within reasonable time and resource constraints. Therefore, we paused the simulation for Ta (Z=21) at 75~fs, as the most energetic electrons in the simulations of both bare Ta (Z=4) and Ta with plastic coating appear by 70~fs.

In addition, the absorption differs by only about 2--3\% between Z=4 and Z=21 for bare Ta, which aligns with our expectation since a larger number of electrons leads to more energy being consumed in ionization processes rather than being absorbed. The relatively small difference also suggests that limiting the simulation to the first four ionization states of Ta may still provide an acceptable quantitative approximation of the overall absorption behavior, despite Ta having up to 73 possible ionization states.

For Ta (Z=4) with plastic coating, the total energy absorbed by all particle species is about 38.5\% of the pulse energy. Beyond the energy absorbed by C and H atoms and their electrons, the absorption only by Ta ions and electrons reaches approximately 28.8\%, still nearly double that of bare Ta. The ion absorption is about 0.6\%, 0.9\%, and 1.0\% for bare Ta (Z=21), bare Ta (Z=4), and Ta with plastic,  indicating that most of the laser energy is absorbed by electrons in these simulations. The absorption efficiency starts to decrease after about 80~fs, as those most energetic electrons leave the simulation box.


\begin{figure}[hbt!]
\centering
\fbox{\includegraphics[height=.22\textwidth]{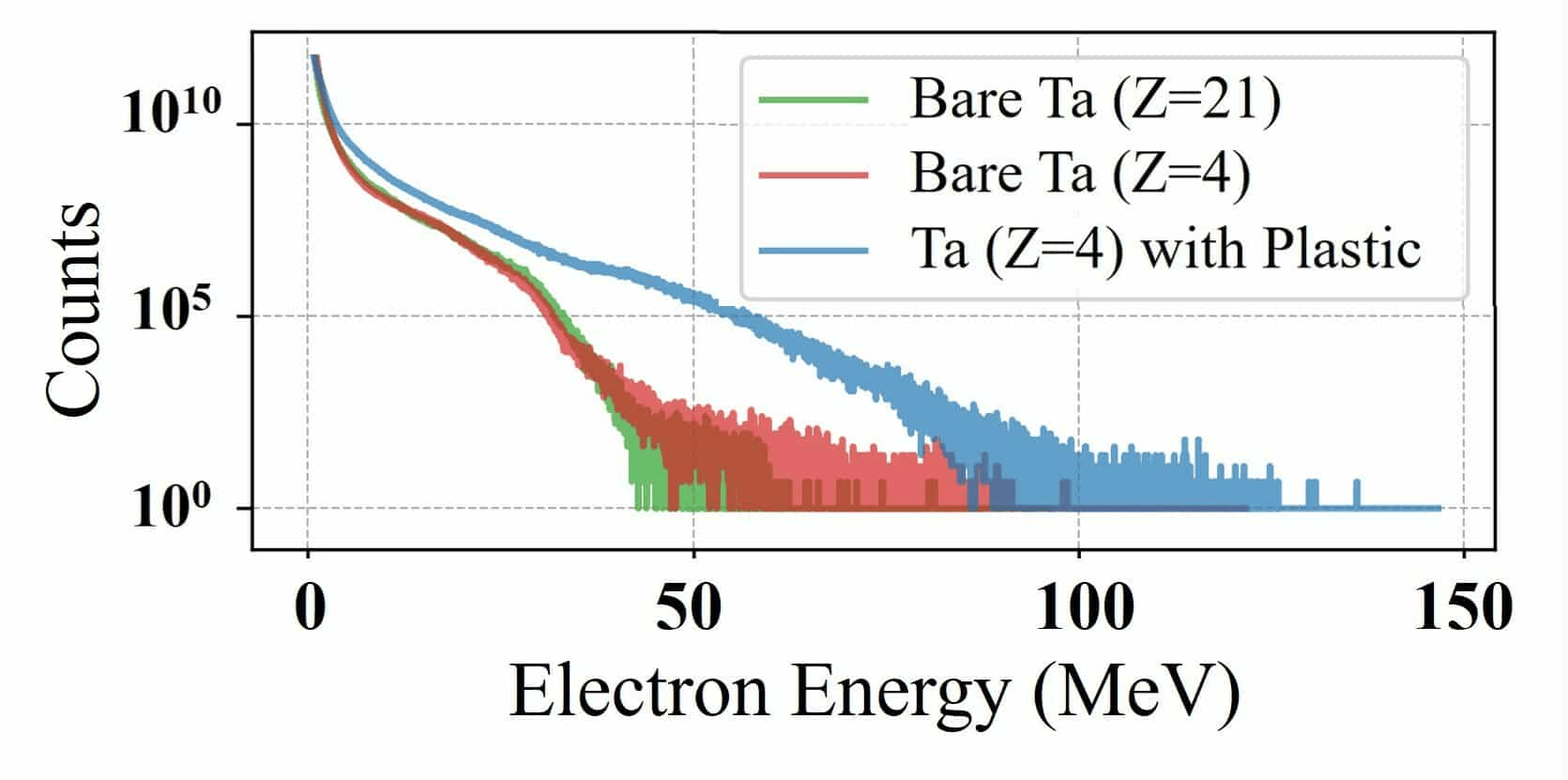}}
\caption{Ta electron spectrum for bare Ta(Z=21), bare Ta(Z=4), and Ta(Z=4) with Plastic coating at 70 fs (when most energetic electrons are observed). In the experimental electron spectra, the bare and plastic-coated Ta produce similar spectra past 30 MeV, with bare Ta generating more electron acceleration.}
\label{fig:Sim_spectrum_ele}
\end{figure}

In Fig.~\ref{fig:Sim_spectrum_ele}, the Ta electron energy spectra at 70~fs for the three cases are presented. Below approximately 60~MeV, the spectra of bare Ta (Z=21) and bare Ta (Z=4) overlap closely. However, for energies above 60~MeV, more electrons gain higher energies in the Z=4 Ta case, with the maximum energy reaching about 98~MeV, compared to approximately 82~MeV for the Z=21 case. A turning point is observed around 30~MeV for both cases.
For Ta with plastic coating, a generally higher number of electrons gain energy across the entire range, with maximum energy reaching about 136~MeV, indicating an enhancement in energy absorption due to the presence of the plastic layer.


\begin{figure}[hbt!]
\centering
\fbox{\includegraphics[height=.22\textwidth]{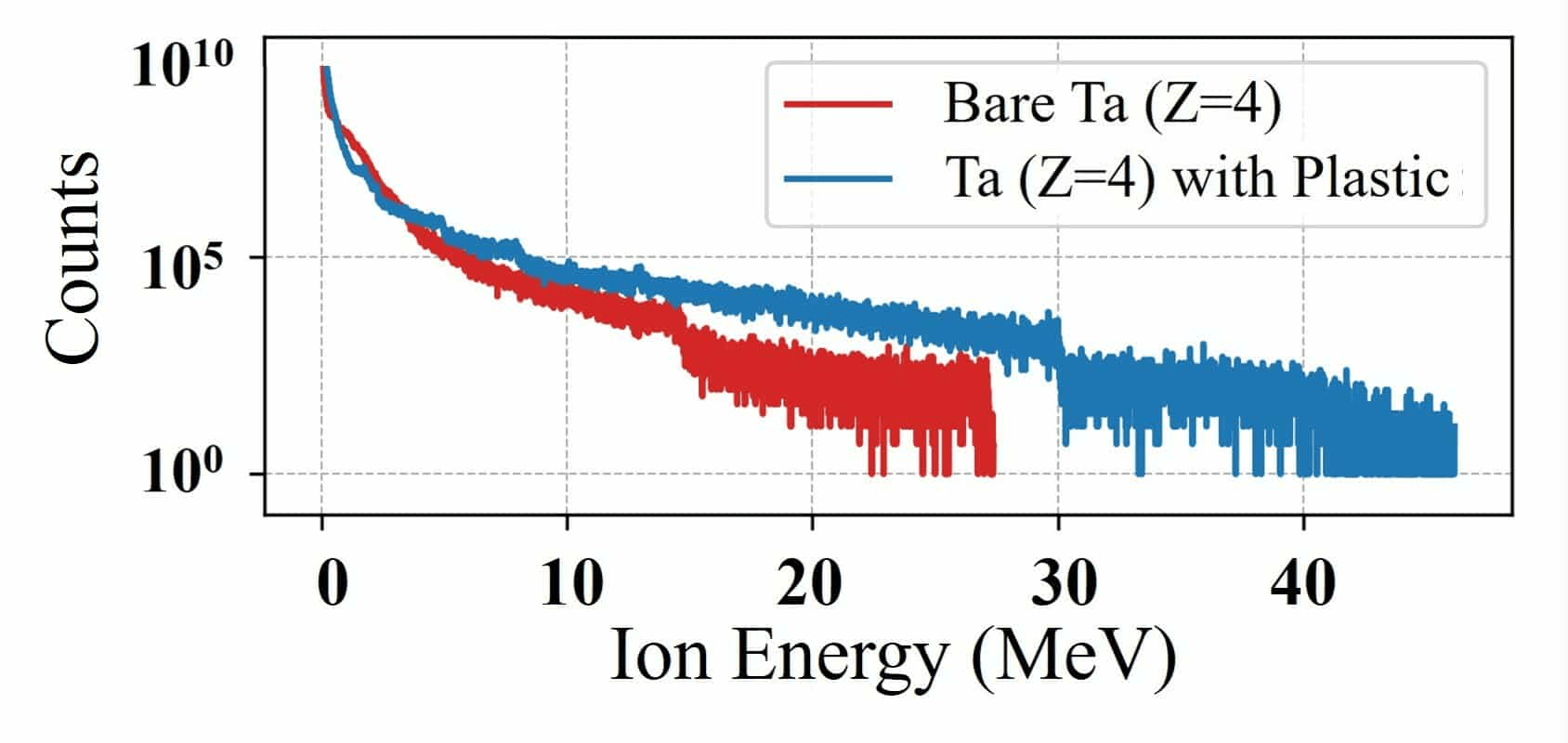}}
\caption{Ta ion spectrum for bare Ta(Z=4), and Ta(Z=4) with Plastic coating at 125 fs (when most energetic ions are observed).}
\label{fig:Sim_spectrum_ion}
\end{figure}

In Fig.~\ref{fig:Sim_spectrum_ion}, the energy spectra of Ta ions with a $4+$ charge state at 125~fs are presented, corresponding to the time when the most energetic ions are observed. It can be seen that, with the plastic coating, there are more ions across the entire energy range above approximately 3~MeV, with the maximum ion energy reaching about 46~MeV. In contrast, for bare Ta, the ion energy extends only up to around 27~MeV. After approximately 150~fs, the maximum ion energies stabilize at around 15~MeV for bare Ta and 30~MeV for Ta with plastic. The energetic electrons transfer energy to the ions as they are pushed into the target, so that a small quantity of ions gain a great amount of energy in a short time. This causes a sharp rise of the spectrum. This trend is also evident in Fig.~\ref{fig:Sim_spectrum_ion} at 125~fs, where distinct steps appear at 15~MeV and 30~MeV for each case, respectively. These sharp features at these energies smooth out at later times.

\section{Discussion}
\label{sec:disc}
The hot electrons generated by the laser-target interaction lose energy to bremsstrahlung radiation, ionization, and electron refluxing, while the highest energy electrons will escape the target \cite{morris2021highly}. In this experiment, laser-target coupling efficiency can be correlated between the reflected light on the MACOR screen and the post-damage craters. A greater signal on the MACOR screen signifies lower laser-target coupling. At the same time, craters measured after the experiment increase in diameter as the laser coupling increases. Measurements from PIC simulations suggest that bare Ta targets exhibited 15~$\%$ energy absorption.  Greater laser-target coupling also leads to greater MeV electron and X-ray generation (Fig. \ref{fig:eSpectra}). Absorption measurements are difficult to make and may require large diagnostics \cite{eftekhari2022laser}. Crater analysis offers a facile option in determining laser-target energy coupling.

The dominant mechanism for electron acceleration in $\sim$mm thick targets at relativistic intensities is \textit{j$\times$B} \cite{StrehlowThesis2022}. The hot electron temperature which is responsible for accelerating ions, can be approximated by using the ponderomotive energy scaling as \cite{wilks2001energetic}:

\begin{eqnarray}
T_{hot} \approx mc^2
\left(1 + \frac{2U_p}{mc^2}\right)^{1/2}\;,
\\
U_p = 9.33 \times 10^{-14} I \lambda^2    
\label{eq:one}.
\end{eqnarray}

Where \textit{T}\textsubscript{\textit{hot}} is the hot electron temperature, \textit{U}\textsubscript{p} is the ponderomotive potential [eV], and \textit{I} is the peak intensity [W/cm\textsuperscript{2}], and $\lambda$ is the laser wavelength [$\mu$m]. Using this equation for our intensities from 0.5-4$\times$10\textsuperscript{21}~W/cm\textsuperscript{2}, we are able to estimate the electron temperatures from 5.62-15.83~MeV. Derived \textit{T}\textsubscript{hot} from Table \ref{tab:eSpectra} show that the ponderomotive scaling agrees well with the experimentally derived temperature for plastic targets, but underestimates electron temperatures for bare and foam-coated targets. 

Coated (high-Z) targets can offer higher laser-target energy coupling than bare targets; however, the coating should not be prematurely destroyed by near-relativistic intensities \cite{li2022high}. It is likely that the tested foam (or NW) thickness and density are not well suited for these intensities and pulse durations. Yin et al. \cite{yin2024advances} ran multiple simulations regarding foam thickness and density at relativistic intensities, and determined the coating should be chosen such that relativistic induced transparency in the dense plasma coincides with the peak of the pulse, in order to generate the most high energy electrons. It is also observed that too much preplasma can prevent most of the laser energy from interacting with the target \cite{yin2024advances}. Yin. et al. \cite{yin2024advances} tested foam-coated tungsten targets with a BELLA-class laser\cite{leemans2010berkeley}, with similar intensity, 815~nm central wavelength and 300~fs pulse duration, and obtained the greatest electron production with 13~$\mu$m foam thickness, at 27~mg/cc. In our case, we used 50~$\mu$m thick foam with 15~mg/cc. Poor optimization of the NW-coating may also have negatively affected the resultant electron spectra. Parameters such as NW diameter, length, and spacing have an important effect on resultant electron number and temperature \cite{vallieres2021enhanced}; however, in Vallieres et al.\cite{vallieres2021enhanced} all of their tested NW-coated targets performed better than bare foils. 

Electron temperature from plastic-coated targets is slightly lowered compared to bare targets. In theory, plastic-coated targets can more effectively couple the laser energy to the target and can more readily undergo relativistic transparency \cite{strehlow2024mev}. Alternatively, an earlier heater beam can pre-ionize the plastic layer, offering a homogeneous plasma to the main pulse \cite{courtois2009effect}. In these previous studies, a plastic layer of 2 or 10~$\mu$m was used over the high-Z Ta target \cite{courtois2009effect,strehlow2024mev}. In our case, while the electron temperature and counts are slightly lowered, the energy absorption and X-ray emission are comparable to those of bare targets. PIC simulations suggest that a 1~$\mu$m thick plastic coating on Ta could increase coupling up to 28.8\% on the Ta substrate.

Simulations from Miller et al. \cite{miller2023maximizing} show a strong correlation between both laser intensity and plasma scale length in producing the greatest 1-5~MeV X-ray emission. The highest intensity tested of 5$\times$10\textsuperscript{20}~W/cm\textsuperscript{2}, at 25~fs pulse duration, exhibited the greatest electron acceleration >5~MeV, while subsequently emitting the highest X-ray dose \cite{miller2023maximizing}. While coated targets in this experiment have a larger plasma scale length, due to the focusing on the substrate, the maximum intensity at the coating surface is lower for the coated targets (by an order of magnitude for foam targets). Therefore, while bare Ta had the shortest plasma scale length, it also maintained the greatest intensity before the plasma was overcritical. Thinner or lower density coatings, as well as focusing on the surface of the coating, would likely cause all coated targets to outperform bare Ta. 

For MeV X-rays, the difference in performance between the different front-surface targets is apparent. Although the X-ray yield is comparable at lower energies, a difference emerges at \textit{E\textsubscript{photons}} > 5~MeV. Similar to the eWASP measurements, the foam and nanowires suffer at higher energies, speaking to the precise laser-plasma conditions necessary to make these target types advantageous. The foam does outperform other targets from 4-8~MeV, a range that is useful for radiographic purposes, and similarly has a different spectral shape. However, the dominance of bare Ta and plastic-coated Ta at the highest energies is likely due to the enhanced laser-plasma coupling that occurs in the performed plasma. The X-rays benefit from self-focusing of the beam and relativistic induced transparency that allows the beam to penetrate into the overdense plasma, similar to past work \cite{strehlow2024mev}. Previous work by Strehlow et al. \cite{strehlow2024mev} with a 1~$\mu$m, 120~J, 140~fs pulse duration laser obtains 3$\times$10\textsuperscript{-4}~rad~@1m/J sr on bare 1~mm Ta compared to our 4.2$\times$10\textsuperscript{-6}~rad~@1m/J sr.

Target normal sheath acceleration (TNSA) is the dominant ion acceleration mechanism for $\sim$~$\mu$m thick solid targets \cite{wilks2001energetic}, as thicker targets reduce the generated field. In this process, a cloud of electrons in the front and rear of the target accelerate ions on the front and rear side. These ions are accelerated in the normal direction from the target. Ions accelerated from the back of the target generally have higher energies, due to a sharper density gradient than the front \cite{wilks2001energetic,ceccotti2007proton}. The angular divergence of the ions from the front surface should greatly exceed that of the rear surface due to a longer pre-plasma length \cite{wilks2001energetic}. 

The rear CR-39 was in the laser axis direction, and both front and rear CR-39 were a similar distance away from the target (33 versus 31~cm). While TNSA is most effective in accelerating protons and low-Z ions, it is still possible to accelerate mid and high-Z ions \cite{StrehlowThesis2022}, with laser to ion energy conversion efficiency of 4\% reported for heavier ions\cite{borghesi2006fast}. As the accelerated ions in the back come from the back surface, they are assumed to be Ta ions \cite{wilks2001energetic}. Front CR-39 likely detected C and Au ions from the coated targets, but cannot be observed due to saturation. Front ion counts is $\sim$3 orders of magnitude greater than back on the CR-39 (Table \ref{tab:ionSpectra}) but likely contains lower energy particles \cite{wilks2001energetic}. PIC simulations observe accelerated ions of 46~MeV from the plastic-coated Ta, which agrees with experimental results.

Coating a target with nanostructures has lead to greater ion yields \cite{strehlow2022laser}, but is dependent on laser energy, pulse duration, and target thickness due to risk of generation of an overcritical pre-plasma at the incorrect time. This overcritical preplasma may then shutter the main pulse interaction \cite{strehlow2022laser}. More efficient generation of electrons to drive TNSA has been exhibited by a low density coating, like foam \cite{macchi2013ion}. Foam and NW-coated targets exhibit the greatest stability and quantity of accelerated ions for our laser parameters. Lower total hot electron generation but heightened ion acceleration could mean that the foam and NW targets experienced a volumetric effect \cite{almassarani2021parametric}. With the laser focusing on the Ta substrate, it is possible the pulse was prematurely shuttered by these thicker coatings. Almassarani et al. observed a larger spot size can generate more ion acceleration, but lower overall max ion energies \cite{almassarani2021parametric}, albeit in $\sim$~$\mu$m thick targets.

\section{Conclusion}
\label{sec:conc}
Four different target-coating geometries were tested on Scarlet, a high contrast, ultrafast laser. Of the tested coatings, bare targets generated the greatest quantity of electron acceleration, the hottest electrons, as well as the greatest MeV X-ray production. Foam and NW-coated targets, however, did produce the greatest heavy ion acceleration. This improved ion acceleration for structured targets compared to bare Ta, is potentially due to the volumetric effect \cite{almassarani2021parametric}. In previous studies, laser intensity and plasma scale length have been strongly correlated with an increase in MeV electrons and X-rays \cite{miller2023maximizing,simpson2021scaling}. While the plasma scale length for coated targets was increased, this premature shuttering of the rest of the pulse led to a decrease in intensity on the target. 

The greatest MeV electron and X-ray yield has been associated with greater laser-target energy coupling. Images of the MACOR screen reveal that bare Ta and plastic-coated Ta had the greatest laser-target energy absorption. Qualitative measurements about laser absorption from the MACOR agree with post-damage crater diameter. The crater diameter decreased from 1.46 \textpm~0.08~mm, to 0.95 \textpm~0.33~mm from the bare Ta to the NW-coated Ta. Particle-in-cell simulations suggest that bare Ta absorbs 15\% of the laser energy. The shots on NW-coated Ta with the greatest MACOR signal correspond to the smallest diameter craters. Absorption measurements are difficult to carry out and may require a large diffuse reflector or an Ulbricht sphere \cite{eftekhari2022laser}. Post-damage crater analysis is an easy way to compare absorption between target types. Future studies may be able to further study the diameter of craters alongside a diffuse reflector, and compare the absorption measurements to particle-in-cell simulations.


While bare targets perform the best, particle-in-cell simulations suggest that 1~$\mu$m plastic-coating would perform better than bare Ta substrate. Due to the Rayleigh length of the focus, a coating <8~$\mu$m would likely perform better than bare Ta, when focusing the laser on the substrate. The 50~$\mu$m thick foam coating with 15~mg/cc was either too thick or too dense for the focusing geometry. The foam or NW target may have also coupled more laser energy if the laser was focused on the coating. Focusing geometry, laser wavelength, pulse duration, and laser contrast should be carefully considered before selecting a target coating, in order to produce the greatest heavy ion and MeV X-ray emission from specific laser sources. These results showcase particle emission from a high-contrast, short pulse, high-power system, and offer a facile way to benchmark absorption for different targets.

\section*{Supplementary Materials}
\label{sec:supp}
See the supplementary materials for the following: Figure s1, Scanning Electron Microscope (SEM) images of the foam and NW-coated targets. Figure s2, MACOR signal variation and subsequent changes to crater morphology. Figure s3, response matrix used for the FSS unfolding. Figure s4, particle-in-cell simulation setup diagram. Figure s5, electron density and temperature 70~fs after irradiation modeled in particle-in-cell.

\begin{acknowledgments}
Partial funding for shared facilities used in this research was provided by the Center for Emergent Materials: an NSF MRSEC under award number DMR-2011876.
\end{acknowledgments}

\section*{Data Availability Statement}

Data taken and used for this paper is available upon request from the primary author.

\section*{References}
\label{sec:ref}
\nocite{*}
\bibliography{Biblio}

\end{document}